\documentclass[aps,prl,twocolumn, hidelinks,floatfix,longbibliography,showpacs]{revtex4-2}

\usepackage{amsmath,amsfonts,amssymb,bm}
\usepackage{dcolumn}
\usepackage{braket}
\usepackage[final]{graphicx}
\usepackage[update,prepend]{epstopdf}
\usepackage{tabularx}
\usepackage{hyperref}
\usepackage{graphicx}
\usepackage{multirow}
\usepackage{blindtext}
\usepackage{mathtools}

\usepackage{dsfont}
\usepackage{pifont}

\begin{document}
\title{Tracking Chirality in Photoelectron Circular Dichroism}

\author{Marec W. Heger}
\affiliation{Dahlem Center for Complex Quantum Systems and Fachbereich Physik,
             Freie Universität Berlin, Arnimallee 14, D-14195 Berlin, Germany}
\author{Daniel M. Reich}
\affiliation{Dahlem Center for Complex Quantum Systems and Fachbereich Physik,
             Freie Universität Berlin, Arnimallee 14, D-14195 Berlin, Germany}
\email[]{danreich@zedat.fu-berlin.de}

\date{\today}

\begin{abstract}
In photoelectron circular dichroism (PECD) it is generally difficult to trace how and when the chirality of the molecule is imprinted onto the photoelectron. We present simulations of PECD in a simple model and employ chirality measures to establish a quantitative connection between the chirality of the potential, the electronic wave function's chirality, and the anisotropy of the photoelectron distribution. We show that these measures are suitable indicators for chirality, paving the way for tracking the chiral evolution from the nuclear scaffold to the final observable.
\end{abstract}

\maketitle

Chiral molecules play an important role in biology, chemistry, and physics \cite{RitchieJChemPhys1974, JohnstoneJACS2014,CombyPhysChemLett2016,DevesonNatureCom2019,JanssenPhysChemChemPhys2014} since their handedness prominently affects interactions with chiral light and matter. A chiral nuclear scaffold gives rise to a chiral potential and imprints a notion of handedness on the system which can be probed and measured via suitable observables. Prominent examples for such observables are microwave three-wave mixing \cite{LeibscherSymmetry2022,SunJPCLett2023,PattersonNature2013, ShubertAngewChemIntEd2014, LobsigerJPhysChemLett2015, LeePhysRevLett2022,SinghAngewChemIntEd2023,DomingosChemSci2020}, circular dichroism (CD) \cite{LogeChemPhysChem2011,StenerJChemPhys2006,RizzoJChamPhys2008,PulmChemPhys1997,LogeChemPhysLetters2007,BerovaWiley2000}, and photoelectron circular dichroism (PECD) \cite{PowisAdvChemPhys2008,LuxAngewChemIntEd2012,KastnerChemPhysChem2016,CireasaNatPhys2015,FaccialaPhysRevX2023,KastnerJChemPhys2017,BeaulieuNatPhys2018,RaneckyPhysChemChemPhys2022,CombyPhysChemChemPhys2023}, with the latter allowing for large anisotropies in many molecular species.
While it has been shown that PECD in the gas phase is strongly affected by short-range interactions \cite{ArtemyevJChemPhys2022}, it was recently demonstrated that even in a hydrogen atom, excitations to chiral intermediate states lead to a PECD signal despite the absence of a chiral short range potential \cite{MayerPhysRevLett2022}. This highlights the complex interplay of the nuclear geometry, the initial electronic wave function, and the interaction with the circularly polarized driving field.

In general, the initial state inherits its chirality from the nuclear scaffold, which underlines the importance of geometric properties for generating chiral signatures.
Candidates for geometric measures to quantify chirality were previously suggested \cite{GilatJPhysA1989,MayerPhysRevLett2022,NeufeldPhysRevA2020,NeufeldPhysRevA2022, AbrahamJChemPhys2024,GrimmeChemPhysLett1998,GuoChemPhysChem2024} for, e.g., electron wave functions and densities, point mass distributions, photoelectron distributions as well as electric fields. However, the application of such measures as a tool to elucidate the connection between the chirality of the nuclear scaffold, respectively the electronic wave function, and experimental chiral observables has so far been missing.
In this work, we employ chirality measures for potentials and wave functions. We analyze their ability to track the emergence of PECD as a paradigmatic chiral observables and investigate in how far such measures can be used to predict the strength of PECD. To keep the chiral system in our study as simple and tuneable as possible, we consider single-photon ionization of hydrogen subject to an artificial chiral potential. This potential mimics the local chiral environment of electrons in PECD experiments. Time-dependent simulations of the electron dynamics demonstrate a qualitative and quantitative link between the anisotropy in the photoelectron distribution and the chirality measures for the potential and the electronic wave function. 

We employ the Hamiltonian
\begin{equation}
	H(t)=H_\text{H}+gV_\text{chiral}+\boldsymbol{A}(t)\cdot\boldsymbol{p}\,,
	\label{Eq:full_ham}
\end{equation}
with $H_\text{H}$ the Hamiltonian of atomic hydrogen, $V_\text{chiral}$ an artificial chiral potential with a scaling factor $g$, and ${\boldsymbol{A}(t)\cdot\boldsymbol{p}}$ the interaction with a laser field in the electric dipole approximation using velocity gauge.
The potential $V_\text{chiral}$ introduces chirality to the system. We expand it in real-valued spherical harmonics $Y_{lm}$, i.e.,
\begin{equation}
	V_\text{chiral}=V\text(r)\sum_{lm}c_{lm}Y_{lm}(\varphi,\theta)\,,
	\label{Eq:chiral_potential}
\end{equation}
with expansion coefficients $c_{lm}$. This simple model allows us to systematically tune the system's chirality. Our study is focused on the angular dependence which plays the critical role in breaking inversion symmetry. To this end, we treat the radial dependence $V\text(r)$ as a simple prefactor for the spherical harmonic expansion, thus limiting any chirality to emerge from the chiral potential via the $c_{lm}$. $V\text(r)$ is chosen as a screened Coulomb potential with a screening length of $0.21$ nm, localizing the effect of the potential around the center. This screening together with the hydrogenic nature of our system is reminiscent of tightly-bound K shell electrons in a chiral molecule's chiral center. 

We study different potentials $V_\text{chiral}$ by truncating the expansion in $l$ according to Eq.~\eqref{Eq:chiral_potential} at different values. We call this truncation threshold $L$. A restriction to low values of $L$ allows us to construct chiral potentials via a simple spherical harmonics expansion featuring only few coefficients. A similar low-order expansion for states has been applied to study propensity rules in PECD \cite{OrdonezPhysRevA2019}. 
Although in realistic chiral molecules much higher orders in $L$ will commonly be involved \cite{KutscherPhysRevA2023}, electromagnetic potentials with low $L$ can be artificially engineered via generating low-order multipole moments using static electric fields \cite{HigginsPhysRevRes2021}. When imprinting chirality in such an external fashion, one could mimic random orientations of molecules in the gas phase by rotating the fields with respect to the ionizing pulse. Note, however, that the radial dependence of the potential in such a framework would differ from the screened Coulomb profile in our model. 

In photoelectron circular dichroism the system's chirality is probed via ionization using a circularly polarized light field, cf.~the final term in Eq.~\eqref{Eq:full_ham}. This leads to a forwards-backwards asymmetry of photoelectrons even for randomly oriented molecular targets. For simplicity, we restrict ourselves to single-photon ionization in this study. To simulate the PECD we solve the time-dependent Schrödinger equation (TDSE) utilizing the Runge-Kutta propagator. We represent the angular degree of freedom of the electronic wave function via spherical harmonics and the radial degree of freedom via a Gauss-Lobatto Finite Element Discrete Variable Representation \cite{BalzerPhysRevA2010,RescignoPhysRevA2000}. The photoelectron spectra are continuously extracted during time evolution using the time-dependent surface flux method (t-SURFF) \cite{ScrinziNewJPhys2012, MosertComputPhysCommun2016}. Furthermore, we employ a complex absorbing potential (CAP) \cite{Manolopoulos2002} to avoid unphysical reflections. 
To simulate PECD in the gas phase we need to account for arbitrary orientations of the chiral potential with respect to the propagation direction of the laser field. We calculate the photoelectron spectra of multiple individual orientations $P^{\alpha\beta\gamma}$ obtained by rotating $V_\text{chiral}$ via the Euler angles $\alpha\beta\gamma$. Since every orientation requires a full propagation of the laser-induced dynamics, we use the efficient Lebedev scheme \cite{EdenKMagnReson1998} to keep the numerical effort low. 

To quantify and track how the chirality of the molecular potential is reflected in the electronic wave function we employ a chirality measure expanding the scope of previous suggestions~\cite{GilatJPhysA1989,MayerPhysRevLett2022,NeufeldPhysRevA2020,NeufeldPhysRevA2022,AbrahamJChemPhys2024}. We emphasize that any chirality measure should ensure that achiral objects are mapped to a value of zero and chiral objects to a nonzero value. To this end, we need to distinguish the quantification of chirality for potentials and wave functions. For the case of potentials - or, more generally, real-valued scalar functions $f(\boldsymbol{r})$ - we define
\begin{align}
	\chi^\text{V}(f(\boldsymbol{r}))&=\frac{\min_{\alpha\beta\gamma}{ \int_{\mathbb{R}^3}\lvert R^{\alpha\beta\gamma}Pf^c(\boldsymbol{r})-f^c(\boldsymbol{r})}\rvert^2d^3r}{4\int_{\mathbb{R}^3}\lvert f(\boldsymbol{r})\rvert^2d^3r}\,.
	\label{Eq:chi_pot}
\end{align}
The idea of Eq.~\eqref{Eq:chi_pot} is to find the minimal overlap between $f^c(\boldsymbol{r})$ and its mirror image $Pf^c(\boldsymbol{r})$ via arbitrary rotations $R^{\alpha\beta\gamma}$ in a proper frame of reference. To perform the rotations in such a frame we define $f^c(\boldsymbol{r})=Tf(\boldsymbol{r})$ with $T$ a translation such that the first moment of $|f^c(\boldsymbol{r})|^2$ is zero.
It is critical to evaluate the minimization over all rotations in a suitable frame of reference. This is because a chirality measure which only minimizes over all rotations in an arbitrary frame can lead to achiral objects being assigned a non-zero value. One can avoid this issue altogether by performing a minimization over all translations on top of the rotations \cite{GilatJPhysA1989}, however, this is impractical. An alternative approach \cite{AbrahamJChemPhys2024} is to evaluate the measures with minimization only over rotations in a coordinate system where the first moment, i.e.~the ``center of mass'', of $\lvert f(\boldsymbol{r}) \rvert^2$ vanishes. In the Supplementary Material we prove that in such a frame of reference the chirality measure from Eq.~\eqref{Eq:chi_pot} is guaranteed to yield $\chi=0$ for achiral objects even when only minimizing over the set of rotations. Our proof also demonstrates that chiral objects are guaranteed to yield nonzero chirality measure. Moreover, the resulting value of $\chi$ will always be an upper bound to a chirality measure which includes minimization over both rotations and translations.

Equation \eqref{Eq:chi_pot} cannot be used as a chirality measure for wave functions. This is because states which differ only by a global phase would not be treated equivalently by the above definition. To amend this, we represent the state as a density matrix to ensure that any global phase has no impact on the chirality measure,
\begin{align}
	\chi^{\rho}(\rho)&=\frac{\min_{\alpha\beta\gamma}\lVert R^{\alpha\beta\gamma}P\rho^c P^\dagger R^{\alpha\beta\gamma\dagger}-\rho^c\rVert^2_\text{HS}}{2\lVert \rho\rVert^2_\text{HS}}\,.
	\label{Eq:chi_rho}
\end{align}
Here, $\lVert\cdot \rVert_\text{HS}$ is the Hilbert-Schmidt norm. Following an analogous approach to Eq.~\eqref{Eq:chi_pot}, a proper frame of reference is chosen by employing $\rho^c$ which is obtained from the density matrix $\rho$ via a translation to a frame in which the first moment, i.e.~the position expectation value, is zero. This guarantees that $\chi^\rho$ is zero if and only if the state is an achiral object. The measures from Eqs.~\eqref{Eq:chi_pot} and \eqref{Eq:chi_rho} are normalized such that they take values in the interval $[0,1]$. We highlight that both measures are scalar - not pseudoscalar - quantities. This avoids the occurrence of chiral zeros \cite{WeinbergCanJChem2000}. This comes at the cost of not being able to characterize the handedness of an object, since, e.g., two enantiomeric forms of a chiral object are mapped to the same value.

\begin{figure}
	\includegraphics[width=\linewidth]{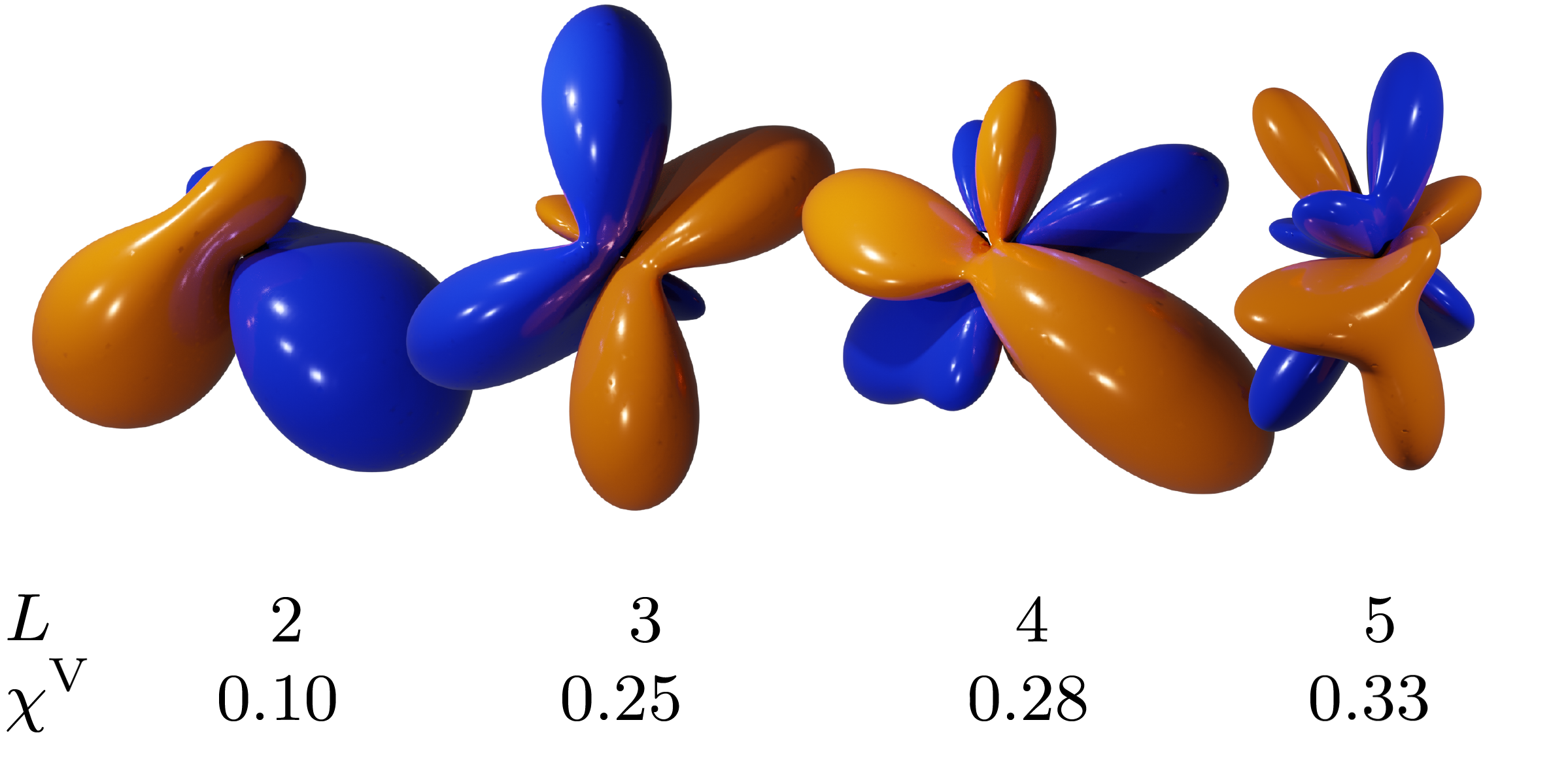}
    \caption{Isosurfaces with values -0.22 (blue) and 0.22 (orange) for chiral potentials according to Eq.~\eqref{Eq:chi_pot} with optimized coefficients $c_{lm}$ for angular expansions up to $L=2-5$ and $g=0.747$. The value of the chirality measure according to Eq.~\eqref{Eq:chi_pot} is indicated below each potential.}
	\label{Fig:chiral_potentials}
\end{figure}

\begin{figure}
	\includegraphics[width=\linewidth]{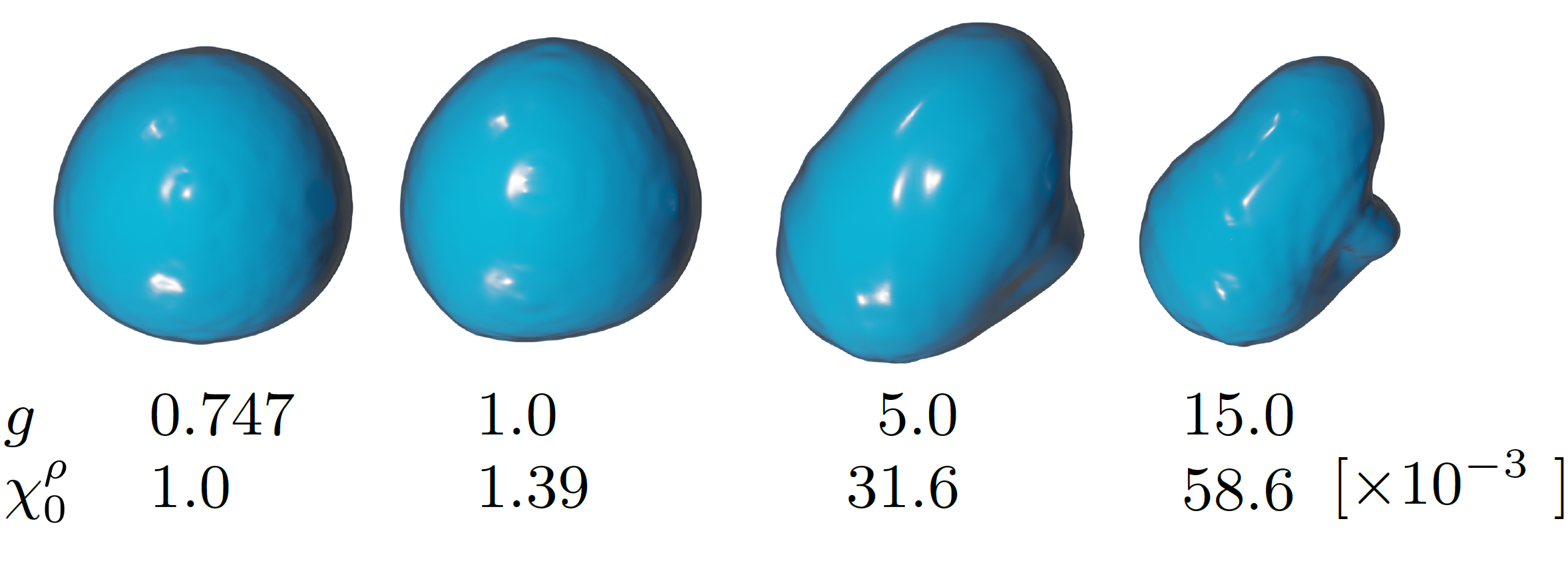}
    \caption{Isosurfaces with value 0.03 of the ground-state electron density for the optimized $L=3$ chiral potential and different strengths $g$. The value of the chirality measure according to Eq.~\eqref{Eq:chi_rho} is shown below each state.}
	\label{Fig:groundstate_potentials}
\end{figure}

Recently, it was shown that optimizing the nuclear scaffold in a model system allows to maximize PECD \cite{RudorffJChemPhys2024}.
Instead of directly using the observable as the figure of merit, we consider in our simulations optimized angular chiral potentials with respect to the geometric notion of chirality as quantified by the chirality measure $\chi^{V}$ from Eq.~\eqref{Eq:chiral_potential} (see Supplementary Material for details). The potentials we obtain are thus optimized independently of a particular observable yet still allow us to ascertain how the inherent chirality of the molecular structure is reflected in PECD. Our results are visualized in Fig.~\ref{Fig:chiral_potentials}. We observe that for a larger truncation limit $L$, the attainable values of $\chi^{V}$ increase as well. This is because a larger value of $L$ allows for more parameters and thus degrees of freedom which can be explored during the optimization. 

The potential imprints chirality onto the electron, which is often prominently reflected in the ground state wave function $\ket{\psi_0}$. The connection between potential and ground state is particularly direct for small chiral strengths $g$ where the chiral potential can be considered as a perturbation to the achiral $H_{\mathrm{H}}$, cf.~Eq.~\eqref{Eq:full_ham}. In the Supplementary Material we show via perturbation theory that to leading order this leads to a chirality measure of the ground state $\chi^\rho_0$ which is proportional to $g^2$. In the perturbative regime, increasing $g$ thus directly translates to scaling the chirality of the system. Figure \ref{Fig:groundstate_potentials} visualizes the gradual deformation of the ground-state electron density for the $L=3$ chiral potential for increasing $g$, illustrating how the chirality measure for states, Eq.~\eqref{Eq:chi_rho}, captures this increase. For small $g$ the ground state is predominantly of $s$-type, with increasing contributions from $p$ and $d$ orbitals for larger values of $g$. For increasing values of $g$ the electron density starts to follow the attractive (blue) part of the chiral potential, cf.~Fig.~\ref{Fig:chiral_potentials}. In all simulations of PECD throughout this work we ensured to remain in the regime where the chiral potential can be seen as a perturbation. This is a realistic assumption for, e.g., the local chiral potentials of K-shell electrons in chiral molecules which inspired our model.

\begin{figure}
	\includegraphics[width=\linewidth]{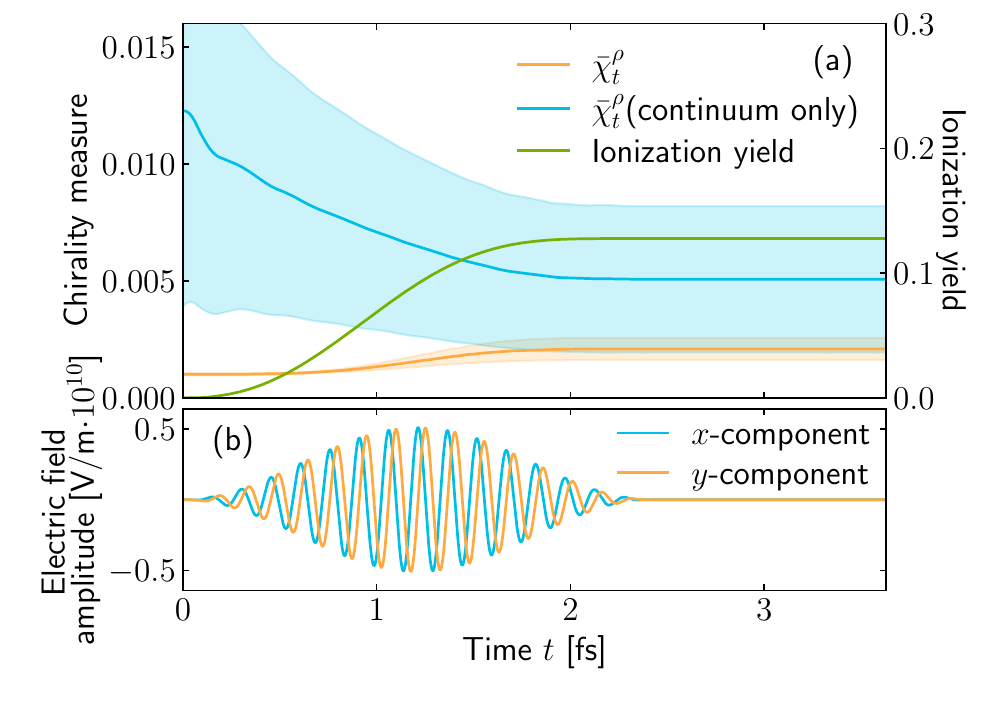}
	\makeatletter\long\def\@ifdim#1#2#3{#2}\makeatother
    \caption{(a) Orientationally averaged chirality measure of the time-evolved state (orange) and for only the continuum part of the electronic wave function (blue) for the optimized $L=3$ potential with $g=0.747$, and total ionization yield (green). The light-orange shaded area shows the 1-$\sigma$ deviation of the chirality measure for the time-evolved state among all orientations. The light-blue shaded area analogously shows the 1-$\sigma$ deviation for the continuum state. (b) Electric field components of the circularly polarized pulse in the polarization plane.}
	\label{Fig:chirality_increase_over_time}
\end{figure}

We employ the chirality measure to track the dynamic evolution of chirality. To this end, we average the chirality measure for the electronic wave function over all orientations at time $t$, denoting the rotationally averaged chirality measure $\bar\chi^\rho_t$.
Figure \ref{Fig:chirality_increase_over_time} shows $\bar\chi^\rho_t$ for the time-dependent wave function using the $L=3$ chiral potential with $g=0.747$ (cf.~Fig.~\ref{Fig:chiral_potentials}). The ionizing pulse is circularly polarized with a duration of $2.41$ fs, a wavelength of $45.5$ nm, and a sine-squared temporal envelope with an amplitude of $5\cdot 10^9$~V/m. The ionization yield obtained by this pulse is shown in Fig.~\ref{Fig:chirality_increase_over_time}(a). The 1-$\sigma$ standard deviation for $\bar\chi^\rho_t$ over all orientations is indicated by the orange-shaded area.  Furthermore we show the orientaionally averaged chirality measure of the continuum part of the electronic wave function only (blue). This continuum part is obtained via projection of the total wave packet onto the set of field-free energy eigenstates, obtained by diagonalizing $H_\text{H} + g V_\text{chiral}$, and subsequent removal of all wave packet components with negative eigenvalue. Analogously, the light-blue shaded area displays the 1-$\sigma$ standard deviation for the continuum electron. The time-dependent cartesian components of the pulse in the polarization plane are shown in Fig.~\ref{Fig:chirality_increase_over_time}(b). We find that the chirality measure of the time-dependent electronic wave function steadily increases during the ionization, reaching a final-time value larger than the chirality measure of the ground state. This trend is not limited to the $L=3$ potential but can be observed for all optimized potentials we studied in the perturbative regime (data not shown). The chirality measure for the continuum part of the electronic wave function shows a large onset chirality which steadily decreases as the electron moves out of the chiral potential. The value at final time is, however, still larger than the chirality of the total wave function, highlighting the role of the continuum in manifesting the chiral signature. We attribute the large initial values for the chirality measure to the fact that the electron still remains close to the chiral center and thus inside the chiral potential at early times. An appreciable part of this chirality is lost once the electron propagates outward such that it can be interpreted as ``transient chirality''. For large times, only the non-transient chirality remains, as tracked by the convergence of our chirality measure to an asymptotic value and it is only this chirality which is captured in the photoelectron angular distribution and thus the PECD signal. Details on our analysis of the transient chirality are provided in the Supplementary Material.

We also investigate whether chirality of the ground state and of the electronic wave function can be related to the strength of the chiral observable, i.e., the PECD signal.
The PECD signal can be characterized by expanding the orientationally averaged photoelectron angular distribution via odd Legendre coefficients $c_i$ which gives rise to the so-called linear PECD (LPECD) \cite{LuxChemPhysChem2015}. The $c_i$ are generally energy-dependent - we consider their values at the ionization peak. Since we study single-photon ionization, only $c_1$ contributes to the expansion \cite{MansonRevModPhys1982}, i.e.,
\begin{equation}
	\text{LPECD}=\frac{1}{c_0}\left(2c_1-\frac{1}{2}c_3+\frac{1}{4}c_5+\dots\right)\xrightarrow{\text{1 ph.}}\frac{2c_1}{c_0}\,,
	\label{Eq:LPECD}
\end{equation}
with $c_0$ the total ionization signal. Note that in order to ensure a fair comparison we tuned the pulse frequency such that the photoelectron peak position consistently stays at $13.6$\,eV for all chiral potentials. We provide further details in the Supplementary Material.
To separate the emergence of chirality in the ionization process from the chirality of the initial ground state, we adjust the strength of the chiral potential $g$ in all simulations such that the chirality measure of the ground state is kept at a constant value of $10^{-3}$. Concretely, this requires $g$ to be increased for potentials with higher $L$ cutoff. We attribute this to the fact that spherical harmonics with higher $l$ have a diminished influence on the ground state due to the larger energetic distance from the ground state. This leads to a reduced imprint of the chiral potential onto the ground state for these higher-order terms.

\begin{figure}
	\includegraphics[width=\linewidth]{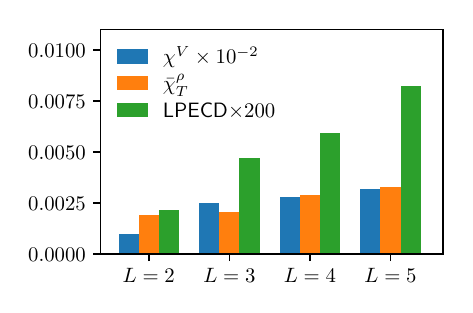}
	\caption{Chirality measure for potential, wave function at final time, and LPECD for optimized potentials with different value of $L$, cf.~Fig.~\ref{Fig:chiral_potentials}. The potential scaling factor $g$ is chosen such that $\chi^\rho_0=10^{-3}$ throughout.}
	\label{Fig:chiLPECD_comparison}
\end{figure}

Figure \ref{Fig:chiLPECD_comparison} shows the orientationally averaged chirality measure of the wave function at final time $\bar\chi^\rho_T$, the chirality measure of the potential $\chi^V$, and the value of the resulting LPECD. Strikingly, Fig.~\ref{Fig:chiLPECD_comparison} demonstrates that all chiral quantities move in tandem with increasing $L$. This is a strong indicator for the predictiveness of the chirality measures for both the potential and the final state in terms of the experimentally observable chiral signature. In addition to the results shown in Fig.~\ref{Fig:chiLPECD_comparison} we performed further simulations adjusting the pulse frequency to shift the photoelectron peak to 10.9\,eV, 20.4\,eV and 27.2\,eV. All calculations displayed the same qualitatively and quantitatively behavior discussed previously, i.e.~a steady increase in LPECD for potentials with higher $L$ with larger chirality measure for the states at final time. Thus, our results extend to a broad range of photoelectron energies. Nevertheless, we expect that the predictiveness of our chirality measures will have some limits. For example, in our model the chiral potential acts only perturbatively on both bound and continuum states of a hydrogenic system. This leads to a ground-state with strong $s$-type character. Since we consider a single-photon ionization process, the most important part of the continuum are states with predominant $p$ character. If a chiral potential has large coefficients for $l=1$ this can potentially lead to a higher "effective" chirality for such states which is not distinguished from higher $l$ contributions by the measure. Similarly, although PECD from excited states goes beyond the scope of this work, ionization from other initial states featuring, e.g., $p$- or $d$-type character will likely alter the effect of the chiral potential on the electron appreciably for example by affecting the population of individual partial waves in the continuum. Even for the ionization from the ground state studied in this work, we found that for potentials with higher $L$ a trend towards populating higher partial waves in the continuum can be observed. As a result, we attribute the larger values of the chirality measure for the electronic state at final time at least partially to higher angular momentum contributions, similarly to the increased chirality measures for the potential which can be attained with larger truncation thresholds $L$.

In conclusion, we have shown in a simple model that chirality measures for potentials and wave functions can be used as a tool to track the emergence and predict the strength of chiral observables, such as PECD. Specifically, a large chirality measure of the potential directly translates into a large chirality measure for the ground states and eventually a large PECD signal in our simulations. We emphasize that chirality measures need to be adapted to the physical problem. For example, to guarantee that achiral objects are assigned a chirality measure of zero it is important to account for all rotations \emph{and} translations or at least to move to a proper frame of reference.
Further development of the chirality measures towards fully taking into account translations will be subject to future work. 
Although no single measure of chirality can be expected to be a perfect quantitative predictor for arbitrary chiral observables, the performance of the chirality measures in the examples we studied here is encouraging in terms of their potential applicability in a wider range of systems and for other chiral observables like, e.g., circular dichroism.

\begin{acknowledgments}
We would like to thank Christiane Koch, Raoul Ebeling, Alexander Blech, and Bar Ezra for helpful discussions. Financial support by the Deutsche Forschungsgemeinschaft (DFG, German Research Foundation)—Projektnummer 328961117—SFB ELCH 1319 is gratefully acknowledged.
\end{acknowledgments}

%

\end{document}


\fontsize{9pt}{16pt}\selectfont
\renewcommand{\baselinestretch}{0.8}
\selectfont
\title{Supplemental Material:\\ Tracking Chirality in Photoelectron Circular Dichroism}

\author{}
\affiliation{}

\maketitle

In the following we provide further details on the properties of the chirality measure discussed in the main text. In particular, we expand on the role of translations, discuss how to account for them, and provide numerical guidance to evaluate the chirality measure in a single-center expansion. We furthermore provide the coefficients for the optimized chiral potentials we employed in our simulations and discuss our results for time-dependent chirality measures for photoelectrons in greater detail, particularly regarding our observations for the continuum part of the electronic wave function. Finally, we show a brief calculation regarding the relationship between the chirality of the ground state and the strength of the chiral potential and provide further details on our numerical simulations and our evaluation of the LPECD in them.

\subsection{Properties of Chirality Measures}

A completely faithful chirality measure for a real-valued function $f(\boldsymbol{r})$ with $\int_{\mathbb{R}^3}\lvert f(\boldsymbol{r})\rvert^2d^3r=1$ needs to account for minimization over all possible translations $T_{xyz}$ in addition to rotations. This is because an object is only chiral if it cannot be superimposed via \emph{translations and rotations} with its mirror image. For quantifying chirality of such scalar functions the corresponding chirality measure reads as follows,
\begin{align}
	\chi^\text{V}\textbf{(}f(\boldsymbol{r})\textbf{)}&=\frac{1}{4}\min_{\stackrel{\alpha\beta\gamma}{xyz}}{ \int_{\mathbb{R}^3}\lvert T^{xyz}R^{\alpha\beta\gamma}Pf(\boldsymbol{r})-f(\boldsymbol{r})}\rvert^2d^3r\,\nonumber\\
	&=\frac{1}{2}-\frac{1}{2}\max_{\stackrel{\alpha\beta\gamma}{xyz}}\int_{\mathbb{R}^3}f(\boldsymbol{r})T^{xyz}R^{\alpha\beta\gamma}Pf(\boldsymbol{r})d^3r\,,
	\label{Eq:chi_pot_sup}
\end{align}
where $\alpha, \beta, \gamma$ are the Euler angles for the rotation operator $R^{\alpha\beta\gamma}$, and $P$ is the operator of spatial inversion. The second line in Eq.~\eqref{Eq:chi_pot_sup} is obtained by expanding the $L^2$ norm and using the fact that all involved operators are unitary. 

For quantifying chirality of a state we need to account for the irrelevance of the global phase and obtain the following, similar expression. For simplicity we assume that $\rho=\ket{\psi}\bra{\psi}$ is a pure state with $\braket{\psi|\psi}=1$. A generalization to mixed states is straightforward,
\begin{align}
	\chi^{\rho}\textbf{(}\rho\textbf{)}&=\frac{1}{2}\min_{\stackrel{\alpha\beta\gamma}{xyz}}\lVert T^{xyz}R^{\alpha\beta\gamma}P\rho P^\dagger R^{\alpha\beta\gamma\dagger}T^{xyz\dagger}-\rho\rVert^2_\text{HS}\,\nonumber\\
	&=1-\max_{\stackrel{\alpha\beta\gamma}{xyz}}\lvert \bra{\psi}T^{xyz}R^{\alpha\beta\gamma}P\ket{\psi}\rvert^2\,.
	\label{Eq:chi_rho_sup}
\end{align}
Note that any global phase obtained by applying $T^{xyz}R^{\alpha\beta\gamma}P$ to $\ket{\psi}$ is absorbed due to the modulus. We normalized our measures such that they take values between 0 and 1. 

While Eqs.~\eqref{Eq:chi_pot_sup} and \eqref{Eq:chi_rho_sup} are very faithful to the geometric nature of chirality, determining the extrema over all $\alpha\beta\gamma,xyz$ is often numerically unfeasible. For both potentials and states we use in our hydrogenic model a single-center expansion, thus rotations can be realized via Wigner D-matrices and are comparatively simple to perform. Conversely, translations in a single-center expansion create slowly converging sums over an infinite range of partial waves $l$. 

Neglecting the minimization over translations will always overestimate the chirality measures from Eqs.~\eqref{Eq:chi_pot_sup} and \eqref{Eq:chi_rho_sup} since in this case the minimal value is only searched over a subset of parameters. While obtaining an upper bound for $\chi$ is in itself quite useful, it is in our opinion critical to not lose fundamental properties of the chirality measure, namely (i) achiral objects always yield $\chi=0$ and (ii) chiral objects always yield $\chi>0$. Although the latter requirement is still fulfilled even if translations are neglected, the former condition does not hold in general in this case. To exemplify this, consider an arbitrary achiral and localized object in three-dimensional space very far away from the origin. The mirrored object is then also very far away from the origin, but in opposite direction. If translations are not considered, then the chirality measure seeks the optimal match of the object with its mirror image by rotation only. However, the mirror image has to first be brought spatially close to the original object to obtain any overlap in the first place, which reduces the three degrees of freedom for the rotation to one. This is generally insufficient to map all achiral objects to $\chi=0$.

\subsection{Adjusting the Frame of Reference}

To solve the issue in the example explained above, we show that by adjusting the frame of reference it can be guaranteed that $\chi=0$ is obtained for achiral objects even when minimization only occurs over rotations. The proof proceeds as follows: We begin by assuming $f(\boldsymbol{r})$ to be achiral. This means, that there exists a solution with $\alpha_s\beta_s\gamma_s,x_sy_sz_s$ such that
\begin{equation}
	f(\boldsymbol{r})=T^{x_sy_sz_s}R^{\alpha_s\beta_s\gamma_s}Pf(\boldsymbol{r})\,.
\end{equation}
For any arbitrary functional $G\textbf{(}f(\boldsymbol{r})\textbf{)}$ the relation
\begin{equation}
	G\textbf{(}f(\boldsymbol{r})\textbf{)}-G\textbf{(}T^{x_sy_sz_s}R^{\alpha_s\beta_s\gamma_s}Pf(\boldsymbol{r})\textbf{)}=0
	\label{Eq:functionals}
\end{equation}
must hold. From here we choose 
\begin{equation}
	G\textbf{(}f(\boldsymbol{r})\textbf{)}=\int_{\mathbb{R}^3}\boldsymbol{r}\lvert f(\boldsymbol{r})\rvert^2d^3r\,.
	\label{Eq:specific_functional}
\end{equation}
Equation \eqref{Eq:functionals} allows us to draw the following conclusion: Since $\lvert f(\boldsymbol{r}) \rvert^2$ is normalized and qualifies as a density we can identify Eq.~\eqref{Eq:specific_functional} as the first moment of $\lvert f(\boldsymbol{r}) \rvert^2$. If this first moment coincides with the origin, then rotations and reflections leave its value invariant. In this case, Eq.~\eqref{Eq:functionals} can only be valid if the translation operator is the identity. In the chirality measure for states the first moment refers to
\begin{equation}
	G\textbf{(}\ket{\psi}\textbf{)}=\braket{\psi|\boldsymbol{\mathsf{r}}|\psi}\,,
\end{equation}
with $\boldsymbol{\mathsf{r}}$ the position operator. This shows that in a coordinate system where the first moment is at the origin no translations are required to obtain $\chi=0$ for achiral objects. The proof for the chirality measure for states proceeds analogously. Due to the significantly reduced complexity of evaluating the minimum only over rotations without needing to account for translations, we decided to employ these simplified chirality measures throughout the main text.

\subsection{Evaluation of $\chi$ for Shifted Functions in Single-Center Expansions}

To evaluate the simplified chirality measures from the main text, a single translation, i.e.~a shift of the coordinate system, still needs to be performed. This needs to be carefully addressed in a single-center expansion. We propose to calculate $\chi$ in momentum space where rotations retain their numerical complexity. This is because positions and momenta behave equivalently under rotations. Conversely, translations become phases in momentum space due to the momentum operator being the generator of translations which is diagonal in momentum space. Equation \eqref{Eq:chi_pot_sup}, and analogously Eq.~\eqref{Eq:chi_rho_sup}, highlight that the greatest degree of numerical complexity lies in the evaluation of $\bra{\psi}T^{xyz}R^{\alpha\beta\gamma}P\ket{\psi}$. As discussed before, we simplify the chirality measure by considering $\bra{\bar\psi}R^{\alpha\beta\gamma}P\ket{\bar\psi}$ instead, where $\ket{\bar\psi}=T^{\boldsymbol{a}}\ket{\psi}$ with $T^{\boldsymbol{a}}$ is a translation to a coordinate system where the first moment is at the origin. We focus our discussion on the chirality measure for states in the following, the case for potentials proceeds analogously. First, we write our states in the following basis,
\begin{align}
	\phi_{klm}(\boldsymbol{r})=f_k(r)Y_l^m(\Omega_r)\,,
\end{align}
with radial basis functions $f_k(r)$ and (complex-valued) spherical harmonics $Y^m_l(\Omega_r)$. The Fourier transform to momentum domain then reads,
\begin{align}
	\phi_{klm}(\boldsymbol{p})&=\frac{1}{\sqrt{2\pi}^3}\iiint e^{-i\boldsymbol{p}\cdot\boldsymbol{r}}\phi_{klm}(\boldsymbol{r})d\boldsymbol{r}\nonumber\\
	&=\sqrt{\frac{2}{\pi}}\iiint \sum_{LM}(-i)^L j_L(pr)Y_L^M(\Omega_p)Y^{M*}_{L}(\Omega_r)\nonumber\\
	&\quad\times Y_l^m(\Omega_r)f_k(r)r^2\sin{\theta}drd\phi d\theta\nonumber\\
	&=\sqrt{\frac{2}{\pi}}(-i)^lY_l^m(\Omega_p)\cdot g^k_l(p)\,,
	\label{Eq:momentum_basis}
\end{align}
with
\begin{align}
	g^k_l(p)=\int j_L(pr)f_k(r)r^2dr
	\label{Eq:g_term}
\end{align}
and
\begin{equation}
	e^{-i \boldsymbol{p} \cdot \boldsymbol{r}}=4\pi\sum_l^\infty\sum_m(-i)^l j_l(pr)Y_l^m(\Omega_p)Y_l^{m*}(\Omega_r)\,.
	\label{Eq:plane_wave_expansion}
\end{equation}
Here, $j_l$ denotes the spherical Bessel function of first kind and $\Omega=(\theta$,$\phi$) are the spherical angles. 
$\ket{\psi}$ can then be expanded via coefficients $c_{klm}$, i.e.,
\begin{align}
	\ket{\psi(\boldsymbol{p})}=\sum_{klm}c_{klm}\phi_{klm}(\boldsymbol{p})\,.
	\label{Eq:momentum_state}
\end{align}
The overlap integral from the chirality measure can then be expressed as
\begin{align}
	\bra{\bar\psi}R^{\alpha\beta\gamma}P\ket{\bar\psi}&=\bra{T^{\boldsymbol{a}}\psi(\boldsymbol{p})}R^{\alpha\beta\gamma}P\ket{T^{\boldsymbol{a}}\psi(\boldsymbol{p})}\nonumber\\
	&=\bra{\psi(\boldsymbol{p})}e^{-i\boldsymbol{a}\cdot\boldsymbol{p}}R^{\alpha\beta\gamma}Pe^{i\boldsymbol{a}\cdot\boldsymbol{p}}\ket{\psi(\boldsymbol{p})}\nonumber\\
	&=\bra{\psi(\boldsymbol{p})}e^{-i\boldsymbol{a}\cdot R^{\alpha\beta\gamma\dagger}\boldsymbol{p}-i\boldsymbol{a}\boldsymbol{p}}\ket{\psi(-R^{\alpha\beta\gamma\dagger}\boldsymbol{p})}\nonumber\\
	&=\bra{\psi(\boldsymbol{p})}e^{-i\boldsymbol{p}\cdot R^{\alpha\beta\gamma}\boldsymbol{a}-i\boldsymbol{p}\boldsymbol{a}}R^{\alpha\beta\gamma}P\ket{\psi(\boldsymbol{p})}\,.
\end{align}
In the following we use the abbreviation $\boldsymbol{b} = R^{\alpha\beta\gamma}\boldsymbol{a}-\boldsymbol{a}$. Substituting Eq.~\eqref{Eq:momentum_basis} and Eq.~\eqref{Eq:momentum_state} and employing another plane wave expansion (cf.~Eq.~\eqref{Eq:plane_wave_expansion}) leads to

	\begin{align}	
		\bra{\bar\psi}R^{\alpha\beta\gamma}P\ket{\bar\psi}&=8\sum_{l,l_1,l_2,k_1,k_2}\sum_{m,m_1,m_2}c^*_{k_1,l_1,m_1}c_{k_2,l_2,m_2}(-i)^{l_2-l_1+l}Y^{m*}_{l}(\Omega_b) \times \nonumber\\
		& \qquad\int Y_{l_1}^{m_1*}(\Omega_p)Y_{l}^m(\Omega_p)R^{\alpha\beta\gamma}PY_{l_2}^{m_2}(\Omega_p) \sin{\theta}d\Omega_p\int g^{k_1}_{l_1}(p)^*g^{k_2}_{l_2}(p)j_l(bp)p^2dp\nonumber\\
		&=8\sum_{l,l_1,l_2,k_1,k_2}\sum_{m_1,m_2}c^*_{k_1,l_1,m_1}c_{k_2,l_2,m_2}A_{m_1,m_2}^{l,l_1,l_2}R^{l,l_1,l_2}_{k_1,k_2}\,,
	\end{align}
	with
	\begin{align}
		R^{l,l_1,l_2}_{k_1,k_2}=(-i)^{l_2-l_1+l}\int g^{k_1}_{l_1}(p)^*g_{l_2}^{k_2}(p)j_l(bp)p^2dp
		\label{Eq:R_term}
	\end{align}
	and
	\begin{align}
		A_{m_1,m_2}^{l,l_1,l_2}&=\sum_m Y^{m*}_{l}(\Omega_b)\int Y_{l_1}^{m_1*}(\Omega_p)Y_l^m(\Omega_p)R^{\alpha\beta\gamma}PY_{l_2}^{m_2}(\Omega_p) \sin{\theta}d\Omega_p\nonumber\\
		&=(-1)^{l_2-m_1}\sum_{M} [D_{m_2M}^{l_2}(R)]^*\sqrt{\frac{(2l_1+1)(2l_2+1)(2l+1)}{4\pi}}\left( \begin{array}{ccc}
			l_1&l_2&l\hspace*{-0.1cm} \\                                              
			0&0&0\hspace*{-0.1cm} \\                                            
		\end{array}\right)\nonumber\\
		& \phantom{=} \times \left( \begin{array}{ccc}
			l_1&l_2&l\hspace*{-0.1cm} \\                                              
			m_1&M&-(m_1+M)\hspace*{-0.1cm} \\                                            
		\end{array}\right) \left(Y^{-(m_1+M)}_{l}(\Omega_b)\right)^*.
		\label{Eq:A_term}
	\end{align}
Here, $\left( \begin{array}{ccc}
	\hspace*{0.1cm}\cdot\hspace*{0.2cm}&\cdot\hspace*{0.2cm}&\cdot\hspace*{0.1cm} \\                                              
	\hspace*{0.1cm}\cdot\hspace*{0.2cm}&\cdot\hspace*{0.2cm}&\cdot\hspace*{0.1cm} \\                                            
\end{array}\right)$ are the Wigner 3j-symbols and $D_{mm'}^l$ Wigner D-matrices. The calculation of Eq.~\eqref{Eq:A_term} can be performed as described above, however, evaluating Eq.~\eqref{Eq:R_term} is not quite as straightforward. A suitable approach is to substitute Eq.~\eqref{Eq:g_term} and to switch the order of integration between $p$ and $r$. This allows to evaluate the integrals over three spherical bessel functions analytically \cite{MehremJPhysA2010}. They are given by
	\begin{align}
		i^{l_2-l_1+l}\left( \begin{array}{ccc}
			l_1&l_2&l\hspace*{-0.1cm} \\                                              
			0&0&0\hspace*{-0.1cm} \\                                            
		\end{array}\right)\int j_{l_1}(rp)j_{l_2}(r'p)j_l(bp)p^2dp&=\frac{(-1)^{l_1-l}\pi\beta(\Delta)}{4b}\sqrt{2l+1}\left(\frac{r}{b}\right)^l\left(\frac{r'}{r}\right)^L\sqrt{\binom{2L}{2l}}(2L+1)\times\nonumber\\
		&\quad\left( \begin{array}{ccc}
			l_1&l-L&L'\hspace*{-0.1cm} \\                                              
			0&0&0\hspace*{-0.1cm} \\                                            
		\end{array}\right)\left( \begin{array}{ccc}
			l_2&L&L'\hspace*{-0.1cm} \\                                              
			0&0&0\hspace*{-0.1cm} \\                                            
		\end{array}\right)\left\{ \begin{array}{ccc}
			l_1&l_2&L\hspace*{-0.1cm} \\                                              
			L&l-L&L'\hspace*{-0.1cm} \\                                            
		\end{array}\right\}P_{L'}(\Delta).
	\end{align}
$P_L(x)$ are the legendre polynomials of order $L$, ${\Delta=(r^2+r'^2+b^2)/(2rr')}$, $\beta(\Delta)=\theta(1-\Delta)\theta(1+\Delta)$ with $\theta$ the Heaviside function in half-maximum convention and $\left\{ \begin{array}{ccc}
	\hspace*{0.1cm}\cdot\hspace*{0.2cm}&\cdot\hspace*{0.2cm}&\cdot\hspace*{0.1cm} \\                                              
	\hspace*{0.1cm}\cdot\hspace*{0.2cm}&\cdot\hspace*{0.2cm}&\cdot\hspace*{0.1cm} \\                                            
\end{array}\right\}$ denotes the Wigner 6j-symbols.

\subsection{Optimized Chiral Potentials}
\begin{table}
	\begin{center}
		\begin{tabular}{||c|c|c c c| c c c c c| c c c c c c c| c c c c c c c c c||} 
			\hline
			$c_{lm}$ & l &  & 1 &  & & & 2 &  &  &  & & &3 & &  &  &  &  & & & \,\,\,4 & & &  & \\ [0.5ex] 
			\hline
			& m  & -1& 0 & 1&-2 &-1&0&1&2&-3&-2&-1&0&\,\,\,1&\,\,\,2&\,\,\,3&-4&-3&-2&-1&\,\,\,0&1&2 &3&4\,\,\,\,\,\,\\ [0.5ex] 
			\hline\hline
			L& $\chi$  & & & & &&&&&&&&&&&&&&&&&&&&\\ [0.5ex] 
			2& $\phantom{}$0.10  & & $\sqrt{0.5}$ & &$\frac{3}{5}\sqrt{0.5}$ &&&$\frac{4}{5}\sqrt{0.5}$&&&&&&&&&&&&&&&&&\\ [0.5ex] 
			3& $\phantom{}$0.25  & & $0.5$ & &$-0.5$ &&&&&$\sqrt{\frac{1}{8}}$&&$\sqrt{\frac{3}{8}}$&&&&&&&&&&&&&\\ [0.5ex] 
			4& $0.28$  & & $0.402$ & &$-0.488$ &&&$-0.189$&&$-0.489$&&$0.190$&&&&&$0.149$&&$-0.128$&&&$-0.119$&&$0.485$&\\ [0.5ex] 
			5& $0.33$  & & $0.283$ & &&$0.505$ &&&&$-0.336$&&&&&&&$0.471$&&$0.338$&&&&&&\\ [0.5ex] 
			6& $0.35$  & & $0.242$ & &&$0.540$ &&&&$ 0.187$&&&&&&&$-0.444$&&$-0.108$&&&&&&\\ [0.5ex] 
			\hline
		\end{tabular}
	\end{center}
	\begin{center}
		\begin{tabular}{||c|c| c c c c c c c c c c c| c c c c c c c c c c c c c||} 
			\hline
			$c_{lm}$ & l &  &  &  & & & 5\,\,\,\, &  &  &  & & & & &  &  &  &  &6\,\,\,\, & &  & & &  & \\ [0.5ex] 
			\hline
			& m  & -5& -4 & -3& -2 &-1&0\,\,\,\,\,&1\,\,\,\,\,&2\,\,\,\,\,&3\,\,\,\,\,&4\,\,\,\,\,&5\,\,\,\,\,&-6&-5&-4&-3&-2&-1\,\,\,&0\,\,\,\,&1\,\,\,\,&2\,\,\,\,&3\,\,\,\,&4\,\,\,\,&5\,\,\,&6\,\,\,\,\hspace*{0.2cm}\\ [0.5ex] 
			\hline\hline
			L& $\chi$  & & & & &&&&&&&&&&&&&&&&&&&&\\ [0.5ex] 
			5& $0.33$  & $0.406$ &&$0.167$&&$0.154$&&&&&&&&&&&&&&&&&&&\\ [0.5ex] 
			6& $0.35$  & $-0.458$ &&$-0.216$&&&&&&&&&$-0.275$&&$-0.103$&&$-0.499$&&&&&&&&\\ [0.5ex] 
			\hline
		\end{tabular}
		\caption{Coefficients $c_{lm}$ obtained from optimization towards maximizing chirality with respect to the chirality measure from the main text. Note that all entries which are not explicitly shown correspond to a value of zero. Note that for $L\geq3$ we expect our solutions to only be local extrema.}
		\label{tab:tableCoef}
	\end{center}
\end{table}
In Table~\ref{tab:tableCoef} we report the coefficients $c_{lm}$ we obtained via numerical maximization of the value of the chirality measure among the set of real-valued functions of the form,
\begin{equation}
	f(\boldsymbol{r})=\sum_{l=0}^L\sum_{m=-l}^lc_{lm}Y_{lm}(\Omega_r)\delta(r-1)\,,
	\label{Eq:optimal_potentials_angular}
\end{equation}
where $Y_{lm}$ are real-valued spherical harmonics and the expansion is truncated at order $L$. We focused on only optimizing the angular part and thus fixed the radial dependence of $f(\boldsymbol{r})$ to a $\delta$-function. This leads to the function $f(\boldsymbol{r})$ only taking nonzero values on the surface of a sphere centered around the origin. Note that the radius of the sphere, chosen to be equal to one in Eq.~\eqref{Eq:optimal_potentials_angular}, does not affect the results. In particular, any non-zero translation of the mirror image always yields a vanishing overlap with the original object. For this reason, optimizing the chirality measure for Eq.~\eqref{Eq:optimal_potentials_angular} with respect to rotations only, i.e., neglecting translations, will perfectly reproduce the value of the completely faithful chirality measure from Eq.~\eqref{Eq:chi_pot_sup}.

\subsection{Investigation of Behavior of the Time-Dependent Chirality Measure for the Photoelectron}
In the following we provide a detailed analysis of the decrease of the chirality measure for the continuum part of the electronic wave function shown in Fig.~3 from the main text. We observe that the continuum electron immediately reaches a large chirality measure of around $\bar\chi^\rho_t=0.012$ at the start of the ionization process. Over time, the chirality measure decreases until it converges towards a value of around $\bar\chi^\rho_t=0.005$ for large times.
We attribute this to the fact that during the interaction with the pulse, the field constantly ionizes the electron generating population in a continuum state with comparatively high value of the chirality measure. With increasing time, the electronic wave packet in the continuum travels outwards and leaves the region in which the chiral potential has an appreciable effect which leads to it losing the ``transient chirality'' it possesses inside the chiral potential. At later times, the predominant components of the electronic wave packet in the continuum have lost most of their transient chirality which explains the asymptotically converging behavior of the chirality measure found in Fig.~3 from the main text. 

\begin{figure}
	\includegraphics[width=12.cm]{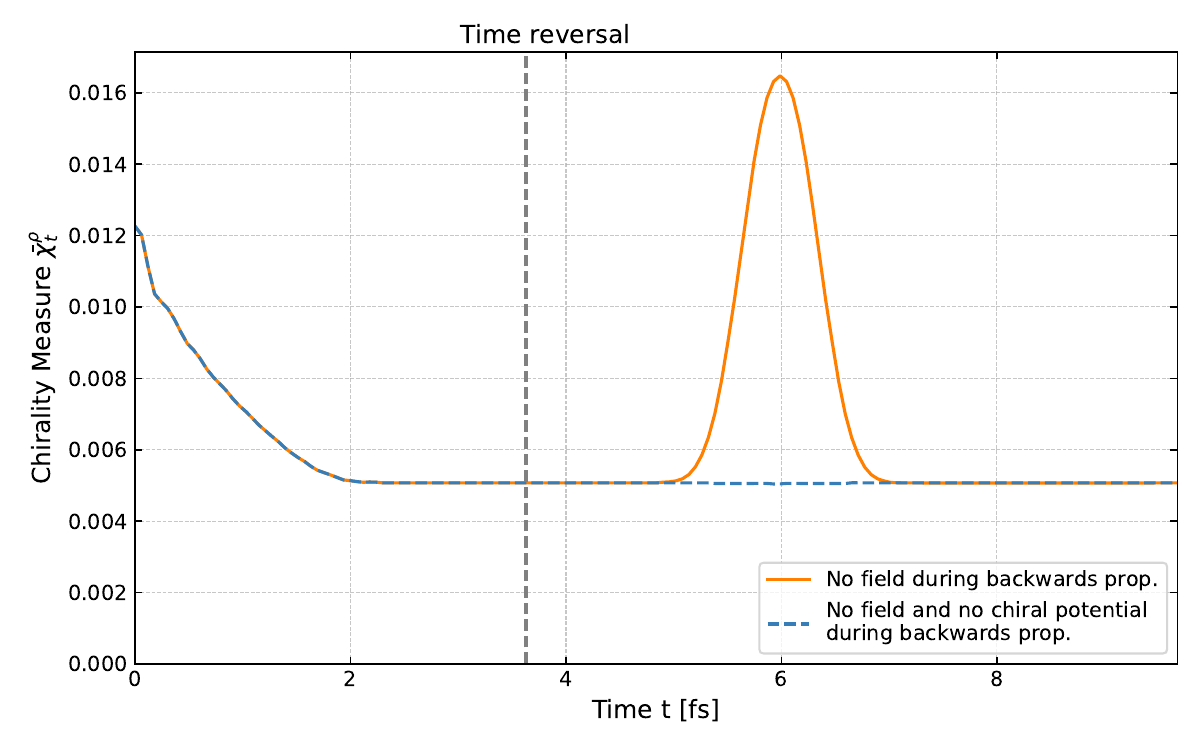}
	\makeatletter\long\def\@ifdim#1#2#3{#2}\makeatother
    \caption{Orientationally averaged chirality measure of the time-evolved state (continuum part only) for the $L=3$ potential with $g=0.747$. At 3.62\,fs the direction of time in the simulation is inverted and the state begins to evolve backwards in time without a driving field. The Hamiltonian for the backward propagation thus contains only the kinetic operator, the Coulomb potential, and the chiral potential (orange), respectively only the kinetic operator and the Coulomb potential (blue). As the electron returns to the core during backwards propagation, the chirality measure increases again indicating transient chirality of the electronic wave function while subjected to the chiral potential. Notably, when the chiral potential is absent during backwards propagation, cf.~the blue curve, no transient chirality is observed.}
	\label{Fig:chirality_plot_time_reversal}
\end{figure}

We performed calculations for the initial dynamics up to 3.62\,fs as before and extended the calculations after this point by continuing the calculations with an inverted flow of time. Furthermore, we removed the circularly polarized pulse and any bound contributions of the electronic wave function at the start of this ``backwards propagation''. The evolution backwards in time leads to the electronic wave packet returning to the nucleus and the chiral potential, scattering with it and then propagating outwards again. The chirality measure for the continuum electron in this setup is shown in Fig.~\ref{Fig:chirality_plot_time_reversal}. While the electron moves back to the center, we observe that the chirality measure increases as the electron reacquires transient chirality while inside the chiral potential, cf.~the orange curve. Since the backwards propagation is performed in the absence of the pulse, this allows us to conclude that the transient increase in chirality in the backwards propagation - or, equivalently, the decrease in chirality in the forward propagation - is not predominantly caused by the circularly polarized field but rather by the chiral potential. To obtain further support for our hypothesis, we performed another simulation in which additionally the chiral potential from the Hamiltonian was removed at the start of backwards propagation at 3.62\,fs. The results are shown as the blue, dotted curve in Fig.~\ref{Fig:chirality_plot_time_reversal}. In the absence of the chiral potential, no appreciable change in the chirality measure can be observed. This allows us to conclude that the initially high value in the chirality measure for the continuum part of the electronic wave function shown in the main text is indeed a transient effect due to the continuum part of the electronic wave function being subjected to the chiral potential.

\subsection{Perturbative Study of Relationship between Chirality of States and Strength of Chiral Potential}

We consider the time-independent Hamiltonian with the chiral potential acting as a perturbation to the hydrogenic Hamiltonian in the regime of small scaling factors $g$ (see Eq.~(1) from the main text). Using perturbation theory, this allows us to derive how chirality from the chiral potential translates to chirality of the ground state as quantified by the chirality measure. The hydrogenic ground state $\ket{E_0^{(0)}}$ of the unperturbed system fulfills
\begin{equation}
	\hat{H}_\text{H}\ket{E_0^{(0)}}=E_0^{(0)}\ket{E_0^{(0)}}\,,
\end{equation}
with $E_0^{(0)}$ the corresponding ground state energy. We denote the remaining set of unperturbed eigenstates via $\{\ket{E_i^{(0)}}\}$  with corresponding eigenenergies $E_i^{(0)}$. Up to second order, the energy shift for the ground state of the chiral system $\ket{E_0^{(2)}}$ is given by time-independent perturbation theory as follows,
\begin{equation}
	E_0^{(2)}=E_0^{(0)}+g\underbrace{\braket{E_0^{(0)}|\hat{V}_\text{chiral}|E_0^{(0)}}}_{= 0}+g^2\sum_{k\neq 0}\frac{\left\vert\braket{E_k^{(0)}|\hat{V}_\text{chiral}|E_0^{(0)}}\right\vert^2}{E_0^{(0)}-E_k^{(0)}}\,.
\end{equation}
Due to the spherical symmetry of the unperturbed ground state the first-order correction vanishes. Thus, the energy shift of the ground state scales with $g^2$ to leading order. The corresponding correction of the ground state to first order reads
\begin{equation}
	\ket{E_0^{(1)}}=\ket{E_0^{(0)}}+g\sum_{k\neq 0}\frac{\braket{E_k^{(0)}|\hat{V}_\text{chiral}|E_0^{(0)}}}{E_0^{(0)}-E_k^{(0)}}\ket{E_k^{(0)}}\,.
\end{equation}
This result allows to to evaluate the chirality measure from Eq.~\eqref{Eq:chi_rho_sup}. We assume that the perturbation does not shift the center of mass of the ground state which is justified for the small values of $g$ considered in a perturbative treatment. This allows to neglect the translation operator. It follows that
\begin{align}
	\chi^{\hat{\rho}}\textbf{(}\ket{E_0^{(1)}}\bra{E_0^{(1)}}\textbf{)}&\approx1-\max_{\alpha\beta\gamma}\lvert \bra{E_0^{(1)}}\hat{R}^{\alpha\beta\gamma}\hat{P}\ket{E_0^{(1)}}\rvert^2\nonumber\\
	&=1-\max_{\alpha\beta\gamma}\Bigg\vert \underbrace{\braket{E_0^{(0)}|\hat{R}^{\alpha\beta\gamma}\hat{P}|E_0^{(0)}}}_{=1}+g^2\underbrace{\sum_{k\neq0}\sum_{k'\neq0}\frac{\braket{E_k^{(0)}|\hat{R}^{\alpha\beta\gamma}\hat{P}|E_{k'}^{(0)}}\braket{E_0^{(0)}|\hat{V}_\text{chiral}|E_{k}^{(0)}}\braket{E_{k'}^{(0)}|\hat{V}_\text{chiral}|E_{0}^{(0)}}}{(E_0^{(0)}-E_k^{(0)})(E_0^{(0)}-E_{k'}^{(0)})}}_{\equiv A(\alpha,\beta,\gamma)}\Bigg\vert^2\nonumber\\
	&=1-\max_{\alpha\beta\gamma}\left\vert1+g^2A\right\vert^2\,,
\end{align}
where we used the orthogonality relation $\braket{E_0^{(0)}|E_k^{(0)}}=\delta_{k,0}$ and the fact that the rotation and parity operator leaves the unperturbed ground state $\ket{E_0^{(0)}}$ invariant. Denoting the angles that maximize the expression above by $\alpha',\beta',\gamma'$, we write the corresponding maximal value of $A$ as $A'\equiv A(\alpha',\beta',\gamma')$. 
The chirality measure then evaluates to
\begin{equation}
	\chi^{\hat{\rho}}\textbf{(}\ket{E_0^{(1)}}\bra{E_0^{(1)}}\textbf{)}\approx1-(1+g^2A')(1+g^2A'^{*})=-4g^2\text{Re}(A')-g^4|A'|^2\,.
\end{equation}
Thus the chirality measure generally becomes proportional to $g^2$ in the perturbative regime to leading order.

\subsection{Further Details on our Numerical Simulations and the Evaluation of LPECD}

All time-dependent simulations in the main text were performed with an adaptive Runge-Kutta propagator of fourth/fifth order. The total propagation time is 3.62\,fs, subdivided into 15000 times steps. The radial grid extends to 300\,Bohr composed of 30 finite elements with 10 grid points each totaling 300 grid points. The angular degree of freedom is expanded into spherical harmonics up to $L=6$. The complex absorber starts at 170\, Bohr and the t-SURFF spectrum is recorded at 150\,Bohr.
To obtain the LPECD in our simulations we proceed as follows. First, we expand the orientationally averaged photoelectron spectra for the enantiomers with handedness $\mu=(+/-)$ in Legendre Polynomials $P_l(\cos\theta)$,
\begin{equation}
\text{PAD}^\mu(k,\theta,\phi)=\sum_lc^\mu_l(k)P_l(\cos\theta)\qquad\text{with }P_{l}(\cos\theta) = \sqrt{\frac{4\pi}{2l+1}}Y_{l}^0(\theta, \phi).
\end{equation}
with coefficients $c^\mu_i(k)$ which depend on the absolute value of the momentum $k$. We now define the quantities
\begin{align}
	c_0(k)&=c_0^-(k)+c_0^+(k), \\
	c_i(k)&=c_i^-(k)-c_i^+(k) \qquad\text{for }i \text{ odd}
\end{align}
which represent the total photon count for both helicities at momentum $k$ and the forward-backwards difference in the basis of Legendre Polynomials. Since our system features predominantly single photon ionization, odd $c_i$ with $i>1$ vanish and we define the LPECD \cite{MansonRevModPhys1982} by
\begin{equation}
	\text{LPECD}(k)\approx\frac{2c_1(k)}{c_0(k)}\,,
	\label{Eq:LPECD}
\end{equation}
where we explicitly denote the dependence on photoelectron momentum compared to Eq.~(5) from the main text for further clarity. We found that the LPECD signal is almost constant around the photoionisation peak in all our simulations. 
Thus we employ its value at $k_\text{peak}=\sqrt{2E_\text{peak}}$ with $E_\text{peak}=13.6$\,eV for simplicity, i.e.
\begin{equation}
	\text{LPECD}\equiv\text{LPECD}(k_\text{peak})\,.
\end{equation}

Table~\ref{tab:L_vals} lists the values $c_0(k_\text{peak})$ and $c_1(k_\text{peak})$ for the LPECD data shown in Fig.~4 of the main text. We observe that the values of $c_0(k_\text{peak})$, corresponding to the total ionization signal, are roughly constant for all chiral potentials we employed. This implies that the differences in LPECD we observe are indeed a consequence of changes to the forward-backwards asymmetry as characterized by $c_1(k_\text{peak})$. 
Furthermore, we provide the orientationally averaged populations in partial waves with angular momenta $l$ for the continuum electronic wave function at final time. 
These shows a general trend towards higher partial waves $l$ for the potentials with larger truncation threshold $L$. Most importantly, we observe that appreciable population in states with up to $l=L+1$ can be found at final time. This can be understood with a simple perturbative argument in which the chiral potential imprints contributions up to $l=L$ on the ground-state which can then be raised further up to $l=L+1$ via ionization by a single photon. As a result, we reason that the higher chirality observed can be at least partially attributed to higher angular momentum contributions, similarly to the higher chirality measures we could reach for potentials with higher $L$.

\begin{table}[h!]
	\centering
	\begin{tabular}{|c|c|c|c|c|}
		\hline
		      & $L=2$ & $L=3$ & $L=4$ & $L=5$ \\ \hline\hline
		$l=0$ & $1.51 \cdot 10^{-6}$ & $5.89 \cdot 10^{-6}$ & $5.70 \cdot 10^{-8}$ & $6.71 \cdot 10^{-8}$ \\ \hline
		$l=1$ & $1.34 \cdot 10^{-1}$ & $1.33 \cdot 10^{-1}$ & $1.33 \cdot 10^{-1}$ & $1.33 \cdot 10^{-1}$ \\ \hline
		$l=2$ & $1.61 \cdot 10^{-3}$ & $2.80 \cdot 10^{-3}$ & $2.59 \cdot 10^{-3}$ & $2.24 \cdot 10^{-3}$ \\ \hline
		$l=3$ & $3.29 \cdot 10^{-5}$ & $4.58 \cdot 10^{-4}$ & $7.02 \cdot 10^{-4}$ & $6.56 \cdot 10^{-4}$ \\ \hline
		$l=4$ & $1.42 \cdot 10^{-8}$ & $7.03 \cdot 10^{-5}$ & $1.59 \cdot 10^{-4}$ & $1.09 \cdot 10^{-4}$ \\ \hline
		$l=5$ & $4.89 \cdot 10^{-9}$ & $3.98 \cdot 10^{-7}$ & $1.90 \cdot 10^{-5}$ & $4.40 \cdot 10^{-5}$ \\ \hline
		$l=6$ & $1.37 \cdot 10^{-10}$ & $4.95 \cdot 10^{-9}$ & $9.27 \cdot 10^{-8}$ & $2.38 \cdot 10^{-6}$ \\ \hline\hline
		$c_0(k_\text{max})$ & $1.384$ & $1.374$ & $1.371$ & $1.368$ \\ \hline
		$c_1(k_\text{max})$ & $2.949 \cdot 10^{-5}$ & $6.446 \cdot 10^{-5}$ & $8.159 \cdot 10^{-5}$ & $1.128 \cdot 10^{-4}$ \\ \hline
	\end{tabular}
    \caption{Orientationally averaged partial wave contributions for final state and coefficients for the LPECD expansion for different values of the truncation threshold $L$ for the chiral potential.}
	\label{tab:L_vals}
\end{table}

%